\documentclass[10pt]{article}
\usepackage[utf8]{inputenc}
\usepackage[T1]{fontenc}
\usepackage{amsmath}
\usepackage{amsthm}
\usepackage{amsfonts}
\usepackage{amssymb}
\usepackage[version=4]{mhchem}
\usepackage{stmaryrd}
\usepackage[colorlinks=true,linkcolor=blue,citecolor=blue,urlcolor=blue]{hyperref}

\newtheorem{lemma}{Lemma}

\newtheorem*{definition*}{Definition}

\newcommand{\kbar}{{\overline{k}}}
\newcommand{\Kbar}{{\overline{K}}}
\newcommand{\tp}{{t^{\prime}}}
\newcommand{\jp}{{j^{\prime}}}
\newcommand{\xp}{{x^{\prime}}}
\newcommand{\xpp}{{x^{\prime \prime}}}
\newcommand{\yp}{{y^{\prime}}}

\newcommand{\ellp}{{\ell^{\prime}}}
\newcommand{\Ip}{{I^{\prime}}}
\newcommand{\Ipp}{{I^{\prime\prime}}}
\newcommand{\maxdisc}{{\rm maxdisc}}
\newcommand{\mindisc}{{\rm mindisc}}
\newcommand{\disc}{{\rm disc}}
\newcommand{\dev}{{\rm dev}}

\newcommand{\yb}{{\bar{y}}}
\newcommand{\xb}{{\bar{x}}}

\title{Whack-a-Mole: Deterministic Packet Spraying Across Multiple Network Paths}

\author{
  Michael Luby$^{1}$ \and John Byers$^{2}$
}

\date{\small
$^{1}$BitRipple, Inc. \quad \href{https://orcid.org/0000-0002-6239-8072}{ORCID: 0000-0002-6239-8072}\\%
$^{2}$Boston University
}

\begin{document}
\maketitle

\section*{Acknowledgments}
The authors gratefully acknowledge support from the U.S. National Science Foundation under
Grant Nos.\ 2212574 (Luby) and 2212575 (Byers).

\begin{abstract}
We present Whack-a-Mole, a deterministic packet spraying algorithm for distributing packets across multiple network paths with provably tight discrepancy bounds. The algorithm is motivated by large-scale distributed AI/ML training and inference workloads, where collective completion time (CCT) and effective training time ratio (ETTR) are highly sensitive to tail latency and transport imbalance. Whack-a-Mole represents the path profile as a discrete allocation of $m$ selection units across $n$ paths and uses a bit-reversal counter to choose a path for each packet. We prove that the discrepancy between expected and actual packet counts per path is bounded by $O(\log m)$ over any contiguous packet sequence. The algorithm responds quickly to congestion feedback by reducing allocations to degraded paths and redistributing load to healthier ones. This combination of deterministic distribution, low per-packet overhead, and compatibility with erasure-coded transport makes Whack-a-Mole an effective building block for multipath transport protocols that aim to minimize CCT and maximize GPU utilization.
\end{abstract}

\section{Introduction}

Large-scale distributed training and inference workloads in AI/ML clusters generate tightly synchronized, high-throughput flows that must complete predictably. Collective operations such as \texttt{AllReduce} and \texttt{AllGather} block compute progress until all participants finish, making \emph{Collective Completion Time (CCT)} and \emph{Effective Training Time Ratio (ETTR)} key metrics for overall job efficiency. Any tail latency or transport imbalance leads to GPU idle cycles, reducing throughput and increasing time-to-train.

Recent internal results (BitRipple LT3, in preparation) demonstrate that ECMP-based flow hashing and retransmission-heavy transport protocols leave significant performance on the table. Queue buildup, packet loss, and Priority Flow Control (PFC) pauses amplify collective latency and create unpredictable completion times. Host-based packet spraying with erasure-coded recovery has been shown to be the only approach to consistently achieve near-optimal CCT across diverse workloads, failure conditions, and buffer sizes. This motivates the need for a packet spraying mechanism that is both efficient and provably well-balanced.

The \emph{Whack-a-Mole} algorithm developed here is designed to be the packet distribution engine that powers BitRipple LT3’s multipath transport. Its purpose is to intelligently and deterministically allocate packets across multiple network paths to maximize aggregate throughput while rapidly adapting to path quality fluctuations. Unlike purely stochastic spraying, which may temporarily overload some paths while underutilizing others, Whack-a-Mole provides low per-packet decision overhead suitable for GPU- or NIC-resident implementation, maintains tightly bounded discrepancy between the desired and achieved path utilization over any time window, and responds quickly to real-time congestion feedback by dynamically ``whacking down'' allocations to degraded paths and redistributing load to healthier ones.

Our design targets several system-level outcomes critical for AI/ML transport: deterministic path utilization so that packet distribution closely follows the target profile even at sub-microsecond packet intervals, graceful adaptation enabling quick recovery when a previously degraded path becomes healthy again, and compatibility with coding-based reliability such as fountain codes or LT3 erasure coding, where only a subset of packets are required for successful decode. 

This paper formalizes a discrete path profile representation, provides algorithms for maintaining and updating these profiles under dynamic network conditions, and derives provable bounds on the deviation between expected and actual packet allocations. These properties make Whack-a-Mole a practical foundation for high-performance, coded, multipath transports such as BitRipple LT3, directly contributing to reduced CCT, improved ETTR, and higher overall GPU utilization.

\section{Motivation}

In packet-switched networks, there are many scenarios where a flow of messages must complete as quickly as possible. Each message consists of a finite set of packets, and a flow refers to the transmission of these packets from a single source to a single destination. Traditionally, message completion is defined as the point at which all packets of a message arrive at the destination. In the setting considered here, we also allow for fountain-encoded messages, where completion occurs as soon as any sufficiently large set of distinct encoded packets, generated from the original message, has arrived and is sufficient for decoding.

To accelerate message completion, packets from a flow may be routed across a set of distinct network paths, which need not be edge-disjoint. Achieving minimal message completion time becomes a load-balancing optimization problem: deciding for each outgoing packet which path to use from the available set. Real networks are highly dynamic. Concurrent flows, link failures, and transient degradation can cause congestion along certain paths, producing increased latency, reduced bandwidth, and packet loss. Different paths may also have inherently different capacities and latencies.

An effective algorithm for path selection must maximize the use of non-degraded paths when they are available, avoid sending packets along degraded paths during congestion events, quickly resume using paths when they recover, and minimize the overall message completion time. This creates a dynamic per-packet decision-making problem at the source, where the goal is to determine how to distribute packets across paths whose performance fluctuates over time.

To guide these decisions, we introduce a \emph{path profile}, which represents how packets should be allocated across the $n$ available paths at a given time. A path profile is an $n$-tuple of nonnegative values
\[
\{p(0), p(1), \ldots, p(n-1)\},
\]
subject to the normalization constraint
\[
1 = \sum_{i=0}^{n-1} p(i),
\]
so that the profile forms a probability density function. Each $p(i)$ specifies the fraction of packets that should be transmitted over path $i$, reflecting the desired load-balancing strategy. The objective is to ensure that approximately a fraction $p(i)$ of packets are routed through path $i$ over time.

One way to implement this objective is to generate a uniformly distributed random number $x \in [0,1]$ for each packet and select the path index $i$ based on $x$ relative to the cumulative distribution function $F$ defined by the profile, choosing $i = F^{-1}(x)$. Although simple, this approach introduces randomness that can lead to short-term variability, especially when the number of packets is small.

For this reason, we focus on discrete rather than continuous representations of path profiles. Discrete representations provide precise control over selection accuracy, avoid inconsistencies from floating-point arithmetic across platforms, deliver predictable behavior over long packet transmissions as profiles evolve, ensure deterministic packet distribution that closely matches the target profile, and require minimal per-packet processing overhead.

Because network conditions vary over time, it is also important to update the path profile dynamically. These updates may reflect static characteristics such as differences in bandwidth or latency between paths, as well as real-time feedback from ECN markings, RTT measurements, or packet loss events. The result is a system where packet spraying decisions continuously track both the desired allocation policy and the evolving state of the network, producing robust and near-optimal message completion times.

\section{Discrete path profiles}

We now describe how to represent a path profile using discrete integers so that packets can be distributed efficiently over paths in accordance with the profile. The representation is designed to be simple to store, easy to update, and fast to use in per-packet decisions.

When there are $n$ available paths, the profile is represented by $n$ bins and $m$ balls, where each bin corresponds to a path and $m$ is the total number of balls. The number of balls $m$ is a measure of precision and is typically fixed for a given system, even though the distribution of balls among bins may change over time as the path profile is updated. The number of balls in bin $i$ is denoted $b(i)$, so that
\[
m = \sum_{i=0}^{n-1} b(i)
\]
is an invariant. The fraction of packets to be sent on path $i$ is then represented by
\[
p(i) = \frac{b(i)}{m},
\]
so that the vector $\{b(0), \ldots, b(n-1)\}$ together with $m$ fully specifies the path profile.

For many operations it is useful to maintain cumulative counts
\[
c(i)=\sum_{j=0}^{i} b(j),
\]
where for convenience we define $c(-1)=0$. The cumulative representation
\[
\{c(-1), c(0), \ldots, c(n-1)\}
\]
is equivalent to the representation using $b(i)$ values, since $b(i)=c(i)-c(i-1)$ for each $i$. The two forms can be computed from each other in linear time. The cumulative form is particularly convenient for path selection: given a packet selection point $k \in \{0,\ldots,m-1\}$, the path chosen for the packet is the smallest $i$ such that
\[
c(i-1) \leq k < c(i).
\]

This discrete integer representation can be stored, updated, and used locally at the source. For example, the source may store $\{b(0),\ldots,b(n-1)\}$ and $m$, or additionally keep $\{c(-1),c(0),\ldots,c(n-1)\}$ precomputed for faster lookups. Other equivalent representations are also possible, and may be chosen to optimize particular operations at the source, but all share the property that they enable efficient packet spraying according to the specified profile with predictable behavior and minimal per-packet processing cost.

\section{Packet spraying across paths}
As described earlier, each bin $i$ can be interpreted as a potential network path $i$, where the value $b(i) / m$ represents the fraction of packets that should be routed along that path. The packet sending method described below deterministically ensures that packets are distributed across the paths as evenly as possible.

Assume that $m$ is a power of two, i.e., $m=2^{\ell}$. An $\ell$-bit spray counter $j$, which increments by 1 modulo $m$ with each packet transmission from a source to a destination, is used to select over which path the current packet is transmitted. 

Define $\theta(j,\ell)$ as the $\ell$-bit integer obtained by reversing the $\ell$ least 
significant bits of $j$ and interpreting it as an integer.  For example, if $\ell=10$ and $j=249$, the
$10$ least significant bits of $j$ are $0011111001$, and the reversed string is $1001111100$, which corresponds to $\theta(j,\ell)=636$.

The path (or bin) selected for the packet with respect to spray counter $j$ is determined by finding the smallest index $i$ such that:

$$
c(i-1) \leq \theta(j,\ell)<c(i)
$$

To determine the appropriate bin $i$, one computes $\theta(j,\ell)$ from the spray counter $j$, and then performs a search over the cumulative distribution array $\{c(-1), c(0), \ldots, c(n-1)\}$. This search can be carried out using linear search, binary search, or interpolation search, depending on the implementation and desired performance characteristics for a given value of $n$.

This packet spraying method offers several advantages: chief among them being its deterministic guarantee that the actual distribution of packets across paths closely adheres to the specified path profile, over any time window.  Examples are described below, which are based on the derivations in Section~\ref{sec: spray bounds}.

The packet spray method has the benefit of being memoryless, in the sense that the path over which to send a packet is based solely on the spray counter and the path profile and does not depend on which paths are used to transmit previous or future packets.

\subsection*{Packet spray counters with shuffling}

In certain scenarios, multiple sources may simultaneously employ packet spraying. If these sources substantially share network paths to their respective destinations and are tightly time-synchronized, they risk generating identical spray counter sequences, which can potentially lead to coordinated transmissions and excessive collisions along shared network paths.

To mitigate this issue, it is advantageous to assign each source a unique spray counter seed, represented as a pair of integers $(sa,sb)$, where
\[
sa \in \{0, 1, 2, \ldots, m-1\}
\]
and
\[
sb \in \{1, 3, 5, 7, \ldots, m-1\},
\]
and thus $a$ is relatively prime to $m$, assuming $m$ is a power of two.

We describe below two packet spray 
methods that use the seed and the packet
sequence number to generate an output that
is then used to choose a path based on the
path profile.  The first method applies
a linear sequence generator to
the packet sequence number and then applies 
a reverse binary operation to produce the
output used to select the path based on the
path profile.  The second
method applies a reverse binary operation
to the packet sequence number and then applies
a linear sequence generator to produce the
output used to select the path based on the
path profile.

Other packet spray methods can be constructed
by taking combinations of these methods. 
For example, two seeds can be used at each
source, where both methods are used 
in combination at the source.

For both methods, the idea is that sources 
use different seeds.  Thus, a source
using a different seed than other sources is likely to choose a different ball than other sources for the same value of $j$. 

The bounds derived in 
Section~\ref{sec: spray bounds} on spray discrepancies remain valid for both of
these enhanced, seed-based packet 
spray counter methods.

To avoid unexpected packet collisions with other sources,  a source may occasionally 
change its seed $(sa,sb)$.  For example, a source could change its seed each time 
$$j \mod m = 0.$$

The path spray counter and seed can be stored and updated 
locally at the source.  Other information based on the
path spray counter and discrete path profile can be stored
and used at the source to optimize processes 
that enable packet spraying across paths based on 
discrete path profiles.

\subsection*{Shuffle method 1}

 For a given source with seed $(sa, sb)$, the spray counter for the $j^{\text{th}}$ packet is computed as follows: the path (or bin) selection is determined by finding the smallest index $i$ satisfying
\[
c(i-1) \leq \theta\left( sa + j \cdot sb, \ell \right) < c(i).
\]

\subsection*{Shuffle method 2}

For a given source with seed $(sa, sb)$, the spray counter for the $j^{\text{th}}$ packet is computed as follows: the path (or bin) selection is determined by finding the smallest index $i$ satisfying
\[
c(i-1) \leq (sa + sb \cdot \theta(j, \ell)) \mod 2^{\ell} < c(i).
\]

\subsection*{Example bounds on spray deviations}

To illustrate the idea of spray deviations, consider a situation where $X$ packets are transmitted between some starting time $t$ and a later time $\tp$. These packets are distributed across different paths according to a path profile, which is represented by a fixed arrangement of $m$ balls, each labeled with a specific path. Now, fix your attention on a particular path, say path $i$.

Based on the path profile, we can determine the expected number of packets that should be transmitted over path $i$ during this interval. This expectation is calculated by looking at the fraction of balls labeled with path $i$ in the path profile and applying that proportion to the total number $X$ of packets transmitted. The result is a real number that represents the idealized or average behavior.

Meanwhile, the actual number of packets transmitted over path $i$ during the same time period is a whole number, since only entire packets are transmitted.

To understand how closely the actual behavior tracks the expected behavior, we define the deviation of path $i$ at start time $t$. This deviation is the smallest range $y$ such that, from time $t$ onward, the actual number of packets transmitted over path $i$ always stays within a moving band of width $y$ around the expected count. More precisely, there must exist a shift $x$, somewhere between $0$ and $y$, so that the actual count is never more than $x$ above the expected count and never more than 
$(y-x)$ below it.

In short, the deviation $y$ captures the tightest tolerance band around the expected transmission pattern that can still fully contain the actual pattern from a given point in time forward.

This is formalized as follows.  By the definition of path profile, the expected number of packets transmitted over path $i$ among $X$ transmitted packets is the real-valued quantity $$B_{i,\left[t, \tp \right]}=\frac{b_{i}}{m} \cdot X.$$ 

Let $A_{i,\left[t, \tp \right]}$ denote the actual (integer-valued) number of packets transmitted over path $i$ out of the $X$ transmitted packets. 
\begin{definition*}
The {\em deviation} of path $i$ starting at time $t$ is the smallest value of $y$ for
which there is an 
$x \in\left[0, y \right]$ such that for all $\tp>t$:
$$
B_{i,\left[t, \tp \right]}+x-y \leq A_{i,\left[t, \tp\right]} \leq B_{i,\left[t, \tp \right]}+x
$$   
\end{definition*}

In Section~\ref{sec: spray bounds} we show
that for all paths $i$ and all start times $t$ the
deviation of path $i$ starting at time $t$
is at most $\ell = \log_2 m$ for shuffle method 1 and at most $2 \cdot \ell$ for
shuffle method 2.

\vspace{.1in}
\noindent
{\bf Example (shuffle method 1)}: Let $m=2^{10}$, i.e., $\ell=10$.
For any $i$ and $t$, 

\begin{itemize}
\item The deviation of path $i$ at start time $t$ is at most $10$.
  \item $\frac{A_{i,\left[t, \tp \right]}}{B_{i,\left[t, \tp \right]}}$ is within an additive amount $10^{-7}$ of $\frac{b_{i}}{\mathrm{~m}}$ for $B_{i,\left[t, \tp \right]} \ge 10^{8}$.
\end{itemize}

Derivations proving these bounds can be found in Section~\ref{sec: spray bounds}.  In practice, discrepancies are often tighter than
the provable bounds.  As an example, $m=1024 = 2^{10}$ and $5$ paths with path profile
$$\left \{\frac{b(0)}{m}, \frac{b(1)}{m},
\frac{b(2)}{m},\frac{b(3)}{m},
\frac{b(4)}{m}\right\}  = 
\left \{ \frac{127}{1024},\frac{400}{1024},
\frac{200}{1024},\frac{173}{1024},
\frac{124}{1024} \right \},$$
the discrepancies for shuffle method 1 starting at time $1$ are 
$$ \{1.9, 1.9, 2.6, 2.5, 2.8\}$$
with respect to a packet spray counter seed $(sa,sb) =(333,735)$.

\section{Packet headers and feedback}
Each packet includes a path identifier in its header, allowing the destination to provide feedback to the source regarding statistics specific to that path.

A path-specific sequence number is included in the packet header, i.e., sequence numbers increment for
packets transmitted over a path independently of other
paths. This design enables the destination to provide detailed feedback to the source on network conditions for individual paths.

For example, if the ECN (Explicit Congestion Notification) bit is set in a packet, the destination can notify the source - using the corresponding path and sequence numbers that the packet experienced congestion. Similarly, the destination can report the round-trip time (RTT) measured for a particular packet, again identified by its sequence and path numbers. Another example includes feedback indicating that, prior to a given packet transmission, other packets transmitted over the same path were lost. The source can use this feedback to refine and update the path profile.

\section{Dynamic adjustment of path profiles}
We describe strategies for updating a path profile. The core idea is to adaptively reduce the allocation to a path when indicators of degraded performance are observed. For example, if packets transmitted over a path exhibit signs such as ECN (Explicit Congestion Notification) markings, increased or above-average one-way delay, or packet loss, the source can reduce the number of packets transmitted along that path.

Specifically, the value of $b(i)$ can be decreased by a constant fraction $\alpha$, where the number of "balls" (representing packet allocation units) removed from path $i$ is:

$$
e(i)=\alpha \cdot b(i)
$$

The whack-a-mole adjustment factor $\alpha$ can vary depending on the severity of the detected issues: smaller values of $\alpha$ are used for minor performance degradations, while larger values are applied for more significant problems.

More generally, the adjustment for path $i$ can be guided by comparing its performance metrics with those of all other paths. The overarching objective is to minimize the overall average severity of issues across all paths, weighted by the relative packet distribution $b(i) / m$.

As a concrete example, each path $i$ can be assigned a severity weight $w(i)$, quantifying the cost or "badness" of using that path. The goal then is to minimize the weighted sum:

$$
\sum_{i=0}^{n-1} w(i) \cdot b(i)
$$

This objective can be achieved by removing allocation (balls) from paths with higher severity weights and redistributing them to lower-cost paths, thereby improving the overall quality of packet delivery.

\section{Updating path profiles}
We describe embodiments where a discrete path profile can be dynamically updated. In a first embodiment of a path profile update, a number $e(j) \leq b(j)$ of balls are removed from bin $j$, and these $e(j)$ balls are as evenly redistributed among the $n$ bins as evenly as possible, with a residual amount distributed round-robin when $n$ does not evenly divide $e(j)$.

To handle the residual balls case, we maintain a (global) residual index $r$ that references the next available residual location.  $r$ is initialized to zero and maintained across updates, as described below.
The function {\tt div} is integer division rounded down, and the function {\tt mod} is the integer modulus function.

\subsection*{Path profile update embodiment 1}
(Remove $e(j)$ balls from bin $j$ and redistribute evenly,
inclusive of bin $j$)

\begin{itemize}
    \item $x=e(j) \operatorname{div} n$
    \item $y=e(j) \bmod n$
    \item Update for each bin $i \neq j$
    \begin{itemize}
        \item $b(i)=b(i)+x$
    \end{itemize}
    \item Update for bin $j$
    \begin{itemize}
        \item $b(j)=b(j)-e(j)+x$
    \end{itemize}
    \item Add back residuals (starting from the last unused residual location $r$)
    \begin{itemize}
        \item For $i=1, \ldots, y$
        \begin{itemize}
            \item $b(r)=b(r)+1 $
            \item $r=(r+1) \bmod n$
        \end{itemize}
    \end{itemize}
\end{itemize}

The first embodiment described above ensures the invariant $m$ on the total number of balls in the bins and spreads the $e(j)$ balls removed from bin $j$ as evenly as possible across all $n$ bins. The "add back residuals" step reestablishes the $m$ balls invariant, which includes the update of the residual index $r$ (global variable) that ensures that bins are equally favored in ball redistribution over the course of multiple path profile updates.

In a second embodiment of a path profile update, balls from multiple bins are removed and redistributed evenly among all bins. Let $\{e(0), e(1), \ldots, e(n-1)\}$ be the number of balls from each of the $n$ bins that are to be removed and redistributed evenly among all the bins, where $e(i) \leq b(i)$ for each bin $k$. As in the first embodiment, a residual index $r$ is maintained across updates to balance residual load fairly.
%

\subsection*{Path profile update embodiment 2}
(Remove from multiple bins and redistribute evenly,
inclusive of bin $j$)

\begin{itemize}
    \item $e=\sum_{i=0}^{n-1} e(i)$
    \item $x=e \operatorname{div} n$
    \item $y=e \bmod n$
    \item Update for bin $i=0, \ldots, n-1$
    \begin{itemize}
        \item $b(i)=b(i)-e(i)+x$
    \end{itemize}
    \item Add back residuals (starting from the last unused residual location $r$)
    \begin{itemize}
        \item For $i=1, \ldots, y$
        \begin{itemize}
            \item $b(r)=b(r)+1 $
            \item $r=(r+1) \bmod n$
        \end{itemize}
    \end{itemize}
\end{itemize}

In a third embodiment of a path profile update, balls are removed from a set of multiple bins and redistributed evenly among the other bins, i.e., those from which no balls were removed.  Let $\{e(0), e(1), \ldots, e(n-1)\}$ be the ball removal profile: the number of balls to be removed from each of the $n$ bins.  We require that $0 \leq e(i) \leq b(i)$ for each bin $i$, so that ball removal is feasible.  We further require that at least one $e(i) > 0$ so that some balls are actually removed, and that at least one $e(j) = 0$, so that removed balls can be redistributed to at least one bin.
We now specify the removal and redistribution process, again with residual replacement.

\subsection*{Path profile update embodiment 3}
(Remove from multiple bins in $K$ and redistribute evenly amongst bins in $\Kbar$)

\begin{itemize}
    \item ${K}=\{i:e(i)>0\}$   
    \item $e=\sum_{i\in K} e(i)$  
    \item $\Kbar =\{i: e(i)=0\}$
    \item $\kbar = |\Kbar|$   
    \item $x=e \operatorname{div} \kbar$
    \item $y=e \bmod \kbar$
    \item Update for bin $i \in K$
    \begin{itemize}
        \item $b(i)=b(i)-e(i)$
    \end{itemize}
    \item Update for bin $i \in \Kbar$
    \begin{itemize}
        \item $b(i)=b(i)+x$
    \end{itemize}
    
    \item Add back residuals, but only to bins in $\Kbar$
    \begin{itemize}
    
        \item while $y > 0$ do
        \begin{itemize}
            \item if $(r \in \Kbar)$ then 
            \begin{itemize}
              \item $b(r) = b(r) + 1$
              \item $y = y - 1$
            \end{itemize}
            \item $r = r + 1 \bmod n$
        \end{itemize}

    \end{itemize}
\end{itemize}

In a fourth embodiment of a path profile update, balls are removed from multiple bins and redistributed proportionally among all bins. In the fourth embodiment, let $\{e(0), e(1), \ldots, e(n-1)\}$ be the number of balls from each of the $n$ bins that are to be removed and redistributed, where $0 \leq e(i) \leq b(i)$ for each bin $i$, and where $k>0$, where $k$ is the number of bins $i$ for which $e(i)=0$.

\subsection*{Path profile update embodiment 4}
(Remove from multiple bins in $K$ and redistribute proportionally across all bins)\\
(Residuals are distributed equally across $\Kbar$)\\
(Recall also that there are $m$ total balls).
\begin{itemize}
    \item ${K}=\{i:e(i)>0\}$   
    \item $e=\sum_{i\in K} e(i)$  
    \item $\Kbar=\{i: e(i)=0\}$
    \item $\kbar = |\Kbar|$   
    \item Update for bin $i=0, \ldots, n-1$
    \begin{itemize}
        \item $b(i)=((b(i)-e(i)) \cdot m) \operatorname{div}(m-e)$
        \item $r(i)=((b(i)-e(i)) \cdot m) \bmod (m-e)$
    \end{itemize}
    \item $\ell=\frac{\sum_{i=0}^{n-1} r(i)}{m-e}$
    \item $x=\ell \operatorname{div} \kbar$
    \item $y=\ell \bmod \kbar$
    \item Update to bins $i \in \Kbar$ (equally):
    \begin{itemize}
        \item $b(i)=b(i)+x$
    \end{itemize}
\item Add back residuals, but only to bins in $\Kbar$
    \begin{itemize}
    
        \item while $y > 0$ do
        \begin{itemize}
            \item if $(r \in \Kbar)$ then 
            \begin{itemize}
              \item $b(r) = b(r) + 1$
              \item $y = y - 1$
            \end{itemize}
            \item $r = r + 1 \bmod n$
        \end{itemize}  
    \end{itemize}  
\end{itemize}

The accompanying Excel spreadsheet provides an example of several of these multiple bin updates for the case when $n=5$ and $m=1024$.

\section{Time-varying path profiles}
In certain scenarios, dynamically updating the path profile over the sending time of a message can significantly improve message completion performance. For example, when some paths have higher latencies than others, it may be advantageous to use those high latency paths only for packets transmitted early in the transmission of a message, and avoid using them for later packets, when doing so could negatively impact timely message completion.

Consider the case of transmitting a 10 Mbit message over two available paths:\\
Path 1: Latency $=100 \mathrm{~ms}$, Bandwidth $=100 \mathrm{Mbps}$\\
Path 2: Latency $=10 \mathrm{~ms}$, Bandwidth $=50 \mathrm{Mbps}$\\
If all packets are transmitted over Path 1 (corresponding to path profile $(1,0)$), the total completion time is $200 \mathrm{~ms}$: $100 \mathrm{~ms}$ to send all data, followed by 100 ms latency for the last packet to arrive. If all packets are transmitted over Path 2 (corresponding to path profile $(0,1)$), the completion time is $210 \mathrm{~ms}$: $200 \mathrm{~ms}$ to send the data and 10 ms latency. If all packets are transmitted at full rate across both paths (corresponding to path profile (0.67,\\
0.33 ), the total completion time is 167 ms: 67 ms to send the data and 100 ms latency on Path 1.

However, a better strategy leverages both paths initially and then switches exclusively to the path with smaller latency:

For the first $37$ ms, send packets at full rate across both paths, using a path profile of (0.67, $0.33)$, i.e., $67 \%$ of packets over Path 1, $33 \%$ over Path 2.

For the remaining $90$ ms, send exclusively over Path 2 using a path profile of $(0,1)$.\\
This hybrid path profile yields a total completion time of 137 ms, faster than any strategy that uses only one path profile during the transmission. This example illustrates how time varying path profiles, tailored to the characteristics of the paths and the transmission timing, can enhance overall performance.

\section{Spray deviation bounds}
\label{sec: spray bounds}

The set of \( m = 2^\ell \) balls can be organized into hierarchical levels of
intervals, indexed by \( e = 0, 1, \ldots, \ell \). At each level \( e \), the
$m$ balls are divided into \( 2^e \) non-overlapping intervals, with each interval
containing exactly \( 2^{\ell - e} \) consecutive balls. 
Each level provides a disjoint and complete covering of the set of $m$ balls.

For example, at level \( 0 \), there is a single interval that includes all \( m \) balls. At level \( 1 \), the balls are divided into two intervals of equal size. More generally, as the level increases, the number of intervals doubles and the length of each interval is halved. Each level refines the previous one: an interval at the 
previous level is partitioned into two equal size intervals at the next level. 
At the most granular level \( \ell \), there are \( 2^\ell = m \) intervals, each consisting of a single ball. 

\begin{definition*}
  The {\em discrepancy} of a set of consecutive
balls $A$ between sequence numbers $j$ 
and $\jp \ge j$, denoted $\disc(A,j,\jp)$, 
is the number of balls chosen from $A$ by the packet spray counter sequence $\{j,\ldots,\jp\}$ minus the expected number $\frac{|A|}{m} \cdot (\jp-j+1)$
of balls chosen from $A$ among $\jp-j+1$ randomly chosen balls. 
\end{definition*}

The discrepancy of $A$ between $j$ and $\jp$  represents the discrepancy between 
how often the packet spray counter actually selects balls from $A$ and how often
we would expect it to, assuming the selection was perfectly uniform at random. 

\begin{definition*}
The {\em maximum discrepancy} of $A$ starting at $j$ is
$$\maxdisc(A,j) = \max_{\jp \ge j} \{0, \disc(A,j,\jp)\}.$$
\end{definition*}

The maximum discrepancy for a set of consecutive balls $A$ starting from 
a certain point $j$, represents the worst-case positive discrepancy between 
how often the packet spray counter actually selects balls from $A$ and how often
we would expect it to, assuming the selection was perfectly uniform at random. 
In other words, it's the greatest amount by which the actual count of 
selections from $A$ exceeds the ideal expected count, across all sequences 
that begin at $j$ and extend forward.

\begin{definition*}
The {\em minimum discrepancy} of $A$ starting at $j$ is
$$\mindisc(A,j) = \min_{\jp \ge j} \{0, \disc(A,j,\jp)\}.$$
\end{definition*}

The minimum discrepancy for a set $A$ starting at position $j$, 
captures the opposite: the worst-case shortfall. 
It’s the greatest amount by which the actual number of times balls 
from $A$ are selected falls short of the expected count, again 
considering all forward sequences beginning at $j$.

\begin{definition*}
  The {\em deviation} of $A$ is 
$$\dev(A) = \max_j \{\maxdisc(A,j)-\mindisc(A,j)\}.$$
\end{definition*}
The deviation of $A$ gives a comprehensive measure of how far the actual behavior of the spray counter
selection can drift from the ideal uniform distribution, in either direction 
or in total spread, with respect to $A$.

The deviation of the level $0$ interval $\{1,\ldots,2^\ell\}$ is always zero (Lemma~\ref{lemma int all}).  
By the way the spray counter operates, 
the deviation of an interval is at most one for shuffle method 1
(Lemma~\ref{lemma int gen 1}) and at most two for shuffle method 2 
(Lemma~\ref{lemma int gen 2}).
A covering of consecutive balls 
$\{ia,\ldots,ib\}$ by (disjoint) intervals is when each ball 
$ic \in \{ia,\ldots,ib\}$ is in exactly one of the intervals.  
By a triangle inequality, the deviation for $\{ia,\ldots,ib\}$ 
is bounded by the minimum number of intervals 
that cover $\{ia,\ldots,ib\}$ times an upper bound on the deviation of an interval.

\begin{lemma}
\label{lemma int all}
For any packet spray counter, for interval $I$ at level $0$,
$$\dev(I) =  0.$$
\end{lemma}

\begin{proof}  
For every $j$, for any $\jp \ge j$, the actual number of balls chosen from $I$ 
by the spray counter is $\jp - j +1$, which is also the expected number
of balls chosen from $I$.
\end{proof}

\begin{lemma}
\label{lemma int gen 1}
Under shuffle method 1, for any interval $I$ at level $e \ge 1$,
$$\dev(I)=1-2^{-e}.$$
\end{lemma}

\begin{proof}
Let $i \in \{0,1\}^e$ identify 
the $(i+1)^{\rm th}$ interval $I$ at level $e$ of size 
$2^{\ell-e}$.
By the definition of $\theta$, the packet spray
counter value $j$ using seed $(sa,sb)$ maps to $I$ if and only if
\begin{equation*}
  \left\lfloor \frac{(  \theta(sa + sb \cdot j,\ell)}{2^{\ell-e}} \right\rfloor = \theta((sa + j \cdot sb ) \mod 2^e, e) =  i.  
\end{equation*}
Since $sb$ and $2^{e}$ are relatively prime and
$\theta(\cdot,e)$ is a one-to-one invertible 
function between $e$-bit integers,
this equation is true for values of $j$
that are separated by $2^e$, i.e., the packet spray counter maps to interval $I$ exactly one out of every $2^e$ consecutive values of $j$.  Since the expected value of mapping to $I$ is $\frac{1}{2^e}$, the expected value goes up by exactly $1$ each $2^e$ balls, and the number of times a ball is actually selected from $I$ is $1$ each $2^e$ balls.  

If the spray counter chooses interval $I$ for value $j$ then 
$$\maxdisc(I,j) = 1-2^{-e}$$ is maximized, in which case $\mindisc(I,j)=0$
and thus $$\maxdisc(I,j)-\mindisc(I,j)=1-2^{-e}.$$

If the spray counter chooses interval $I$ for value $j+2^e-1$ then 
$$\mindisc(I,j) = -(1-2^{-e})$$ is minimized, in which case $\maxdisc(I,j)=0$
and thus $$\maxdisc(I,j)-\mindisc(I,j)=1-2^{-e}.$$

These are the extreme cases for $\maxdisc(I,j)$ and $\mindisc(I,j)$, and for any
other case 
$$\maxdisc(I,j)-\mindisc(I,j) = 1-2^{-e}.$$
From this it follows that $\dev(I) = 1-2^{-e}$. 
\end{proof}

\begin{lemma}
\label{lemma int gen 2}
Under shuffle method 2, for any interval $I$ at level $e \geq 1$,
\[
\dev(I) \leq 2 \cdot (1 - 2^{-e}).
\]
\end{lemma}

\begin{proof}
Let $i \in \{0,1\}^e$ denote the $(i+1)^{\text{th}}$ interval $I$ at level $e$, each of length $2^{\ell - e}$. According to the definition of $\theta$, a packet spray counter value $j$ falls into interval $I$ under seed $(sa, sb)$ if and only if
\begin{equation}
\label{eq:method-2}
\left\lfloor \frac{(sa + sb \cdot \theta(j,\ell)) \bmod 2^\ell}{2^{\ell - e}} \right\rfloor = i.
\end{equation}

For each $x \in \{0,1\}^{\ell - e}$, define the set
\[
J(x) = \{ j = y + x \cdot 2^e : y \in \{0,1\}^e \},
\]
which has size $2^e$. We will show that the left-hand side of Equation~\eqref{eq:method-2} defines a bijection from $J(x)$ to $\{0,1\}^e$.

For any $j = y + x \cdot 2^e \in J(x)$, the value of $\theta(j,\ell)$ is given by
\[
\theta(j,\ell) = \xb + \yb \cdot 2^{\ell - e},
\]
where $\xb \in \{0,1\}^{\ell - e}$ and $\yb \in \{0,1\}^e$ are the reverse binary encodings of $x$ and $y$, respectively. Substituting into Equation~\eqref{eq:method-2}, the numerator becomes:
\[
(sa + sb \cdot \xb + sb \cdot \yb \cdot 2^{\ell - e}) \bmod 2^\ell.
\]
Thus, the expression becomes
\begin{equation}
\label{eq:tech-1}
\left\lfloor \frac{(sa + sb \cdot \xb + sb \cdot \yb \cdot 2^{\ell - e}) \bmod 2^\ell}{2^{\ell - e}} \right\rfloor.
\end{equation}

Let $(sa + sb \cdot \xb) \bmod 2^\ell = \xp \cdot 2^{\ell - e} + \xpp$, where $\xp \in \{0,1\}^e$ 
and $\xpp \in \{0,1\}^{\ell - e}$ are uniquely determined by variable $\xb$ and constants $sa$, $sb$, 
and $\ell$. Then Equation~\eqref{eq:tech-1} simplifies to
\begin{equation}
\label{eq:tech-2}
\left\lfloor \frac{((\xp + sb \cdot \yb) \cdot 2^{\ell - e} + \xpp) \bmod 2^\ell}{2^{\ell - e}} \right\rfloor = (\xp + sb \cdot \yb) \bmod 2^e.
\end{equation}

Since $sb$ is an odd constant and $\xp$ is fixed
by $x$, the map from $J(x)$ to $\{0,1\}^e$ defined
as
$$y + x \cdot 2^e \mapsto (\xp + sb \cdot \yb) \bmod 2^e$$ 
is a bijection. It follows that for any $x$, the counter values in $J(x)$ are mapped evenly across the $2^e$ intervals at level $e$.

Now, consider the deviation for interval $I$. Suppose the spray counter chooses interval $I$ for sequence numbers:
\[
j_1 = x \cdot 2^e - 1,\quad
j_2 = x \cdot 2^e,\quad
j_3 = (x+2) \cdot 2^e - 1,\quad
j_4 = (x+2) \cdot 2^e.
\]
Because
\[
\disc(I,j_1,j_2) = 2 \cdot (1 - 2^{-e}) \quad \text{and} \quad \disc(I,j_1,j_3-1) = 0
\]
and because for every 
$x \in \{0,1\}^{\ell-e}$ the
packet spray counter chooses interval $I$ 
for exactly one sequence number in $J(x)$,
it follows that
\[
\maxdisc(I,j_1) - \mindisc(I,j_1) = 2 \cdot (1 - 2^{-e}).
\]
Because
\[
\disc(I,j_2 + 1,j_3 - 1) = -2 \cdot (1 - 2^{-e}) \quad \text{and} \quad
\disc(I,j_2 + 1,j_4) = 0 \]
and because for every 
$x \in \{0,1\}^{\ell-e}$ the
packet spray counter chooses interval $I$ 
for exactly one sequence number in $J(x)$,
it follows that
\[
\maxdisc(I,j_2 + 1) - \mindisc(I,j_2 + 1) 
= 2 \cdot (1 - 2^{-e}).
\]

These cases represent the maximal span of discrepancy
for interval $I$ given that for every 
$x \in \{0,1\}^{\ell-e}$ the
packet spray counter chooses interval $I$ 
for exactly one sequence number in $J(x)$. 
Therefore, for any starting sequence number $j$, 
the deviation for interval $I$ is bounded by:
\[
\dev(I) \leq 2 \cdot (1 - 2^{-e}).
\]
\end{proof}

\begin{lemma}
\label{lemma sum}
Let $I$ be an interval of size $2^{\ellp}$ and
$ia \in I$, $ib \in I$ where $ia < ib$.
Let $x$ be the minimum number of intervals to cover from
the beginning of $I$ through $ia$, let $y$ be the
minimum number of intervals to cover from $ia+1$ through $ib$,
and let $z$ be the minimum number of intervals to cover
from $ib+1$ through the end of $I$.  Then
when  $\ellp = \ell$ 
the deviation of $\{ia,\ldots,ib\}$ is at most
$$\min\{y, x+z\}$$ and when  $\ellp <\ell$ 
the deviation of $\{ia,\ldots,ib\}$ is at most
$$\min\{y,x+z\}+1$$ 
for shuffle method 1 and at most twice this for shuffle method 2.
\end{lemma}
\begin{proof}
Follows from Lemma~\ref{lemma int all}, 
Lemma~\ref{lemma int gen 1}, 
Lemma~\ref{lemma int gen 2}, 
and a triangle inequality.  
\end{proof}

\begin{lemma}
\label{lemma basic}
Let $I$ be an interval of size $2^{\ellp}$ and
$ia \in I$.
Let $x$ be the minimum number of intervals
(possibly from different levels) it takes to cover
from the beginning of $I$ through $ia$, and let $y$ be the minimum number of intervals it takes 
to cover from $ia+1$ through the end of $I$.
Then, $$x+y \le \ellp+1.$$
\end{lemma}
\begin{proof}
For any integer $ic$ such that 
$0 < ic < 2^{\ellp}$, the total number of 
bits set to one in the binary representations
of $ic$ and $2^{\ellp}-ic$ is $\ellp+1$.
\end{proof}

\begin{lemma}
\label{lemma dev}
   The deviation of $\{ia,\ldots,ib\}$ is at most
   $\ell$ for shuffle method 1 and at most twice this for shuffle method 2.
\end{lemma}
\begin{proof}
Let $\ellp$ be the smallest value such that
$ia$ and $ib$ are in an interval $I$ of 
size $2^{\ellp}$ and $ia$ is in the 
first interval $\Ip$ within $I$ of size $2^{\ellp-1}$ and $ib$ is in the second interval $\Ipp$ within $I$ of size $2^{\ellp-1}$.
Let $x$ be the number of intervals that cover
from the beginning of $\Ip$ through $ia$, 
$y$ be the number of intervals that cover
from $ia+1$ through the end of $\Ip$,
$\xp$ be the number of intervals that cover from
the beginning of $\Ipp$ through $ib$,
and $\yp$ be the number of intervals that cover 
from $ib+1$ through the end of $\Ipp$.  From
Lemma~\ref{lemma basic}, 
$$x+y \le (\ellp-1)+1 = \ellp,$$
$$\xp + \yp \le (\ellp-1)+1 = \ellp,$$
and thus
$$x+y+\xp+\yp \le 2 \cdot \ellp.$$
From Lemma~\ref{lemma int all} and Lemma~\ref{lemma sum}, 
if $\ellp=\ell$ then
the deviation of 
$\{ia,\ldots,ib\}$ is at most
$$\min\{x+\yp,\xp+y\} \le \ell.$$
From Lemma~\ref{lemma int gen 1}, Lemma~\ref{lemma int gen 2}, and Lemma~\ref{lemma sum}, if $\ellp<\ell$ then
the deviation of 
$\{ia,\ldots,ib\}$ is at most
$$\min\{x+\yp,\xp+y\}+1 \le \ellp+1 \le \ell$$
for shuffle method 1 and at most twice this for shuffle method 2.
\end{proof}

\begin{lemma}
The deviation of $\{ia,\ldots,ib\}$ is at most 
$\left \lceil \log_2(ib - ia) \right \rceil +2$ for shuffle method 1 and at most twice this for shuffle method 2. 
\end{lemma}

\begin{proof}
Let 
$\ellp = \left \lceil \log_2(ib - ia) \right \rceil.$
This can be argued based on one of two cases:
\begin{description}
    \item[(1)] There is an interval from level $\ell-\ellp$ of $2^{\ellp}$ balls that starts at or before ball $ia$ and ends at or after ball $ib$.  From the proof of Lemma~\ref{lemma dev}, the deviation of $\{ia,\ldots,ib\}$ is at most $\ellp+1$ for shuffle method 1 and at most twice this for shuffle method 2.
    \item[(2)] There are two intervals, one from level $\ell-\ellp+1$ with $2^{\ellp-1}$ balls and the other from level $\ell-\ellp$ with $2^{\ellp}$ balls, where one of the intervals starts at or before $ia$ and ends before $ib$ and the the other starts where the first interval ends and ends at or after $ib$. From Lemma~\ref{lemma int gen 1}, 
    Lemma~\ref{lemma int gen 2},
    Lemma~\ref{lemma sum}, and Lemma~\ref{lemma basic}
    the deviation of balls $ia$ through $ib$ is at most
    $$ \left \lfloor \frac{(\ellp-1)+1}{2} \right \rfloor
    + \left \lfloor \frac{\ellp+1}{2} \right \rfloor + 2 = \ellp + 2$$
    for shuffle method 1 and at most twice this for shuffle method 2.
\end{description}
Thus, the deviation of $\{ia,\ldots,ib\}$ 
is at most $\ellp+2$ for shuffle method 1 and at most twice this for shuffle method 2.
\end{proof}

\end{document}